\documentclass[3p]{els2article}

\usepackage[utf8]{inputenc}

\usepackage{graphicx}
\usepackage{graphics}
\usepackage{latexsym}
\usepackage{textcomp}
\usepackage{amsmath}
\usepackage{amsfonts}
\usepackage{amssymb}
\usepackage{amsthm}
\usepackage{bussproofs}
\usepackage{stmaryrd}
\usepackage[all]{xy}
\usepackage{cmll}

\usepackage{boxedminipage}
\usepackage{tabularx}
\usepackage{url}
\usepackage{proof}
\usepackage{bussproofs}
\usepackage{latexsym}
\usepackage{stmaryrd}
\usepackage{enumerate}
\usepackage{pdflscape}

\usepackage{multicol}

\def\bs{\boldsymbol}

\def\ra{\rightarrow}

\def\cf{\boxright}

\DeclareSymbolFont{symbolsC}{U}{txsyc}{m}{n}
\DeclareMathSymbol{\strictif}{\mathrel}{symbolsC}{74}
\DeclareMathSymbol{\boxright}{\mathrel}{symbolsC}{128}
\DeclareMathSymbol{\Diamondright}{\mathrel}{symbolsC}{132}
\DeclareMathSymbol{\boxRight}{\mathrel}{symbolsC}{136}
\DeclareMathSymbol{\DiamondRight}{\mathrel}{symbolsC}{140}
\DeclareMathSymbol{\Diamonddot}{\mathrel}{symbolsC}{144}

\newdefinition{definition}{Definition}

\newtheorem{lemma}{Lemma}

\title{PUC-Logic embedding of Lewis' Deontic Logics}

\author[cds]{Ricardo Queiroz de Araujo Fernandes\fnref{fn1}}                                                                            
\ead{ricardo@cds.eb.mil.br}

\author[di]{Edward Hermann Haeusler}                   
\ead{hermann@inf.puc-rio.br}     

\author[df]{Luiz Carlos Pereira}                   
\ead{luiz@inf.puc-rio.br}                                       
                                                                                            
\fntext[fn1]{Thanks to PUC-Rio for the VRac sponsor. Thanks to DAAD (Germany) for the Specialist Literature Programme.}

\address[cds]{Centro de Desenvolvimento de Sistemas\\QG do Exército, Bloco G, 2º Andar\\Setor Militar Urbano, CEP 70630-901\\Brasília, DF, Brasil}

\address[di]{Departamento de Inform\'atica}

\address[df]{Departamento de Filosofia\\Pontif\'icia Universidade Cat\'olica do Rio de Janeiro\\Rua Marquês de São Vicente, 225, Gávea, CEP 22453-900\\Rio de Janeiro, RJ, Brasil}
                                                             
\begin{document}

\begin{abstract} 
We present a embedding of Lewis Deontic logics in PUC-Logic. We achieve this by representing the very basic $\bs{CO}$ logic and showing its relative completeness.
\end{abstract}


\maketitle

\section{Deontic logics}

In \cite{Lewis,LewisPapers}, Lewis presented his approach for Deontic logics based on systems of spheres in comparison to other formalisms. In \cite{Lewis}, he presented two possible definition of the operator $O$, based on his counterfactual operators $\cf$ and $\boxRight$. In \cite{LewisPapers}, he gave the definition of four value structures. The definition based on a nesting $\$$ over the set $I$ is equivalent to the definition of the truth of the operator $\boxRight$. For this reason, we take $O(\phi|\psi) = \psi \boxRight \phi$ as suggested in \cite{Lewis}. This deontic operator can be expressed in terms of labels as follows:
\begin{definition}$$O(\alpha^{\Sigma}/\beta^{\Omega}) \equiv (\beta^{\Omega,\bullet}\wedge(\beta^{\Omega}\ra\alpha^{\Sigma})^{\ast})^{\circledcirc}$$
\begin{center}
$P(\alpha^{\Sigma}/\beta^{\Omega}) \equiv \neg O (\neg(\alpha^{\Sigma})/\beta^{\Omega})$\end{center}
\end{definition} We prove here that the PUC-ND is complete for the $\bs{CO}$-logic according to the axioms and rule of inference below. We write $A$ for $\alpha^{\Sigma}$, $B$ for $\beta^{\Omega}$ and $C$ for $\gamma^{\Theta}$.
\begin{itemize}
\item[R1] All tautologies;
\item[R2] Modus Ponens;
\item[R3] If $A \equiv B$ is theorem, then $O(A/C) \equiv O(B/C)$ is a theorem;
\item[R4] If $B \equiv C$ is theorem, then $O(A/B) \equiv O(A/C)$ is a theorem;
\item[A1] $P(A/C) \equiv \neg O(\neg A/C)$;
\item[A2] $O(A\wedge B/C) \equiv (O(A/C) \wedge O(B/C))$;
\item[A3] $O(A/C) \ra P(A/C)$;
\item[A4] $O(\top_{n}/C) \ra O(C/C)$;
\item[A5] $O(\top_{n}/C) \ra O(\top_{n}/B\vee C)$;
\item[A6] $(O(A/B) \wedge O(A/C)) \ra O(A/B\vee C)$;
\item[A7] $(P(\bot_{n}/C) \wedge O(A/B\vee C)) \ra O(A/B)$;
\item[A8] $(P(B/B\vee C) \wedge O(A/B\vee C)) \ra O(A/B)$.
\end{itemize}

From Fernandes\cite{PUC-Logic-ArXiv}
\begin{lemma} \label{transfer}
Given a theorem $\alpha^{\Sigma}$, there is a proof of $\alpha^{\Sigma}$ in the context $\{N,u\}$, in which the variables $N$ and $u$ do not occur in the proof.
\end{lemma}

We now present a proof for each rule and axiom:\\

\noindent \textbf{(R1)} From completeness of PUC-ND;

\noindent \textbf{(R2)} Modus Ponens is a valid rule in PUC-ND;

\noindent \textbf{(A1)} By definition.\\

\noindent \textbf{(R3)} Given some proof $\Pi \vdash (A\ra B)\wedge(B\ra A)$, by lemma \ref{transfer} and rule 1 of PUC-ND, we have a proof $\Psi \vdash^{N,u}_{N,u} A\ra B$. We present the proof of $O(A/C) \ra O(B/C)$. The proof of $O(B/C) \ra O(A/C)$ is similar.\\

\noindent \textbf{(R4)} Given some proof $\Pi \vdash (B\ra C)\wedge(C\ra B)$, by lemma \ref{transfer}, we have a proof $\Psi \vdash^{N,u}_{N,u} (B\ra C)\wedge(C\ra B)$. We present the proof of $O(A/C) \ra O(A/B)$. The proof of $O(A/B) \ra O(A/C)$ is similar.

\begin{landscape}\begin{center}\AxiomC{$^{3}$[$(C^{\bullet}\wedge(C\ra A)^{\ast})^{\circledcirc}$]}
\UnaryInfC{$(C^{\bullet}\wedge(C\ra A)^{\ast})^{\circledcirc}$} \RightLabel{$\circledcirc$}
\UnaryInfC{$C^{\bullet}\wedge(C\ra A)^{\ast}$}
\AxiomC{$^{3}$[$(C^{\bullet}\wedge(C\ra A)^{\ast})^{\circledcirc}$]}
\AxiomC{$^{2}$[$C^{\bullet}\wedge(C\ra A)^{\ast}$]} \RightLabel{$N$}
\UnaryInfC{$C^{\bullet}\wedge(C\ra A)^{\ast}$} \RightLabel{$N$}
\UnaryInfC{$C^{\bullet}$} \RightLabel{$N$}
\AxiomC{$^{1}$[$C$]} \RightLabel{$N,u$}
\UnaryInfC{$C$} \RightLabel{$N,u$}
\AxiomC{$^{2}$[$C^{\bullet}\wedge(C\ra A)^{\ast}$]} \RightLabel{$N$}
\UnaryInfC{$C^{\bullet}\wedge(C\ra A)^{\ast}$} \RightLabel{$N$}
\UnaryInfC{$(C\ra A)^{\ast}$} \RightLabel{$N,\ast$}
\UnaryInfC{$C\ra A$} \RightLabel{$N,u$}
\UnaryInfC{$C\ra A$} \RightLabel{$N,u$}
\BinaryInfC{$A$} \RightLabel{$N,u$}
\AxiomC{$\Psi$} \RightLabel{$N,u$}
\UnaryInfC{$A\ra B$} \RightLabel{$N,u$}
\BinaryInfC{$B$} \LeftLabel{1} \RightLabel{$N,u$}
\UnaryInfC{$C\ra B$} \RightLabel{$N,\ast$}
\UnaryInfC{$C\ra B$} \RightLabel{$N$}
\UnaryInfC{$(C\ra B)^{\ast}$} \RightLabel{$N$}
\BinaryInfC{$C^{\bullet}\wedge(C\ra B)^{\ast}$} \RightLabel{$\circledcirc$}
\BinaryInfC{$C^{\bullet}\wedge(C\ra B)^{\ast}$}
\UnaryInfC{$(C^{\bullet}\wedge(C\ra B)^{\ast})^{\circledcirc}$} \LeftLabel{2}
\BinaryInfC{$(C^{\bullet}\wedge(C\ra B)^{\ast})^{\circledcirc}$}  \LeftLabel{$\bs{R3}$ \hspace*{1cm} 3}
\UnaryInfC{$(C^{\bullet}\wedge(C\ra A)^{\ast})^{\circledcirc}\ra(C^{\bullet}\wedge(C\ra B)^{\ast})^{\circledcirc}$} \DisplayProof

\AxiomC{$^{2}$[$(C^{\bullet}\wedge(C\ra A)^{\ast})^{\circledcirc}$]}
\UnaryInfC{$(C^{\bullet}\wedge(C\ra A)^{\ast})^{\circledcirc}$} \RightLabel{$\circledcirc$}
\UnaryInfC{$C^{\bullet}\wedge(C\ra A)^{\ast}$}
\AxiomC{$^{2}$[$(C^{\bullet}\wedge(C\ra A)^{\ast})^{\circledcirc}$]}
\AxiomC{$^{1}$[$C^{\bullet}\wedge(C\ra A)^{\ast}$]} \RightLabel{$N$}
\UnaryInfC{$C^{\bullet}\wedge(C\ra A)^{\ast}$} \RightLabel{$N$}
\UnaryInfC{$C^{\bullet}$} \RightLabel{$N,\bullet$}
\UnaryInfC{$C$}
\AxiomC{$^{1}$[$C$]} \RightLabel{$N,u$}
\UnaryInfC{$C$} \RightLabel{$N,u$}
\AxiomC{$\Psi$} \RightLabel{$N,u$}
\UnaryInfC{$(B\ra C)\wedge(C\ra B)$} \RightLabel{$N,u$}
\UnaryInfC{$C\ra B$} \RightLabel{$N,u$}
\BinaryInfC{$B$} \RightLabel{$N,\bullet$}
\UnaryInfC{$B$} \RightLabel{$N$}
\UnaryInfC{$B^{\bullet}$} \LeftLabel{1}
\BinaryInfC{$B^{\bullet}$} \RightLabel{$N$}
\UnaryInfC{$B^{\bullet}$} \RightLabel{$N$}
\AxiomC{$^{1}$[$C^{\bullet}\wedge(C\ra A)^{\ast}$]} \RightLabel{$N$}
\UnaryInfC{$\Pi_{1}$} \RightLabel{$N$}
\UnaryInfC{$(B\ra A)^{\ast}$} \RightLabel{$N$}
\BinaryInfC{$B^{\bullet}\wedge(B\ra A)^{\ast}$} \RightLabel{$\circledcirc$}
\BinaryInfC{$B^{\bullet}\wedge(B\ra A)^{\ast}$} \LeftLabel{1}
\BinaryInfC{$(B^{\bullet}\wedge(B\ra A)^{\ast})^{\circledcirc}$} \LeftLabel{$\bs{R4}$ \hspace*{1cm}2}
\UnaryInfC{$(C^{\bullet}\wedge(C\ra A)^{\ast})^{\circledcirc}\ra(B^{\bullet}\wedge(B\ra A)^{\ast})^{\circledcirc}$} \DisplayProof

$\bs{\Pi_{1}}$\AxiomC{$^{1}$[$B$]} \RightLabel{$N,u$}
\UnaryInfC{$B$}
\AxiomC{$\Psi$} \RightLabel{$N,u$}
\UnaryInfC{$(B\ra C)\wedge(C\ra B)$} \RightLabel{$N,u$}
\UnaryInfC{$(B\ra C)$} \RightLabel{$N,u$}
\BinaryInfC{$C$}
\AxiomC{$C^{\bullet}\wedge(C\ra A)^{\ast}$} \RightLabel{$N$}
\UnaryInfC{$C^{\bullet}\wedge(C\ra A)^{\ast}$} \RightLabel{$N$}
\UnaryInfC{$(C\ra A)^{\ast}$} \RightLabel{$N,\ast$}
\UnaryInfC{$C\ra A$} \RightLabel{$N,u$}
\UnaryInfC{$C\ra A$} \RightLabel{$N,u$}
\BinaryInfC{$A$} \LeftLabel{1} \RightLabel{$N,u$}
\UnaryInfC{$B\ra A$} \RightLabel{$N,\ast$}
\UnaryInfC{$B\ra A$} \RightLabel{$N$}
\UnaryInfC{$(B\ra A)^{\ast}$} \alwaysNoLine \UnaryInfC{$\phantom{-}$} \DisplayProof

\textbf{A2}\AxiomC{$^{2}$[$(C^{\bullet}\wedge(C\ra (A\wedge B))^{\ast})^{\circledcirc}$]}
\UnaryInfC{$(C^{\bullet}\wedge(C\ra (A\wedge B))^{\ast})^{\circledcirc}$} \RightLabel{$\circledcirc$}
\UnaryInfC{$C^{\bullet}\wedge(C\ra (A\wedge B))^{\ast}$}
\AxiomC{$\Pi_{2}$}
\UnaryInfC{$(C^{\bullet}\wedge(C\ra A)^{\ast})^{\circledcirc}$}
\AxiomC{$\Pi_{3}$}
\UnaryInfC{$(C^{\bullet}\wedge(C\ra B)^{\ast})^{\circledcirc}$}
\BinaryInfC{$(C^{\bullet}\wedge(C\ra A)^{\ast})^{\circledcirc} \wedge (C^{\bullet}\wedge(C\ra B)^{\ast})^{\circledcirc}$}
\BinaryInfC{$(C^{\bullet}\wedge(C\ra A)^{\ast})^{\circledcirc} \wedge (C^{\bullet}\wedge(C\ra B)^{\ast})^{\circledcirc}$}
\UnaryInfC{$(C^{\bullet}\wedge(C\ra (A\wedge B))^{\ast})^{\circledcirc} \ra (C^{\bullet}\wedge(C\ra A)^{\ast})^{\circledcirc} \wedge (C^{\bullet}\wedge(C\ra B)^{\ast})^{\circledcirc}$} \alwaysNoLine \UnaryInfC{$\phantom{-}$} \DisplayProof $\Pi_{2}$ and $\Pi_{3}$ are similar.

$\bs{\Pi_{2}}$\AxiomC{$(C^{\bullet}\wedge(C\ra (A\wedge B))^{\ast})^{\circledcirc}$}
\AxiomC{$C^{\bullet}\wedge(C\ra (A\wedge B))^{\ast}$} \RightLabel{$N$}
\UnaryInfC{$C^{\bullet}\wedge(C\ra (A\wedge B))^{\ast}$} \RightLabel{$N$}
\UnaryInfC{$C^{\bullet}$}
\AxiomC{$^{1}$[$C$]} \RightLabel{$N,\ast$}
\UnaryInfC{$C$}
\AxiomC{$C^{\bullet}\wedge(C\ra (A\wedge B))^{\ast}$} \RightLabel{$N$}
\UnaryInfC{$C^{\bullet}\wedge(C\ra (A\wedge B))^{\ast}$} \RightLabel{$N$}
\UnaryInfC{$(C\ra (A\wedge B))^{\ast}$} \RightLabel{$N,\ast$}
\UnaryInfC{$C\ra (A\wedge B)$} \RightLabel{$N,\ast$}
\BinaryInfC{$A\wedge B$} \RightLabel{$N,\ast$}
\UnaryInfC{$A$} \LeftLabel{1} \RightLabel{$N,\ast$}
\UnaryInfC{$C \ra A$} \RightLabel{$N$}
\UnaryInfC{$(C \ra A)^{\ast}$} \RightLabel{$N$}
\BinaryInfC{$C^{\bullet}\wedge (C\ra A)^{\ast}$} \RightLabel{$\circledcirc$} 
\BinaryInfC{$C^{\bullet}\wedge (C\ra A)^{\ast}$}
\UnaryInfC{$(C^{\bullet}\wedge (C\ra A)^{\ast})^{\circledcirc}$} \DisplayProof

\textbf{A2}\AxiomC{[$(C^{\bullet}\wedge(C\ra A)^{\ast})^{\circledcirc} \wedge (C^{\bullet}\wedge(C\ra B)^{\ast})^{\circledcirc}$]}
\UnaryInfC{$(C^{\bullet}\wedge(C\ra A)^{\ast})^{\circledcirc} \wedge (C^{\bullet}\wedge(C\ra B)^{\ast})^{\circledcirc}$}
\UnaryInfC{$(C^{\bullet}\wedge(C\ra A)^{\ast})^{\circledcirc}$} \RightLabel{$\circledcirc$}
\UnaryInfC{$C^{\bullet}\wedge(C\ra A)^{\ast}$}
\AxiomC{[$(C^{\bullet}\wedge(C\ra A)^{\ast})^{\circledcirc} \wedge (C^{\bullet}\wedge(C\ra B)^{\ast})^{\circledcirc}$]}
\UnaryInfC{$(C^{\bullet}\wedge(C\ra A)^{\ast})^{\circledcirc} \wedge (C^{\bullet}\wedge(C\ra B)^{\ast})^{\circledcirc}$}
\UnaryInfC{$(C^{\bullet}\wedge(C\ra B)^{\ast})^{\circledcirc}$}
\UnaryInfC{$C^{\bullet}\wedge(C\ra B)^{\ast}$}
\AxiomC{[$\shpos N$]} \RightLabel{$M$}
\UnaryInfC{$\Pi_{4}$}
\AxiomC{[$\shpos M$]} \RightLabel{$N$}
\UnaryInfC{$\Pi_{5}$}
\BinaryInfC{$(C^{\bullet}\wedge(C\ra (A\wedge B))^{\ast})^{\circledcirc}$}
\BinaryInfC{$(C^{\bullet}\wedge(C\ra (A\wedge B))^{\ast})^{\circledcirc}$}
\BinaryInfC{$(C^{\bullet}\wedge(C\ra (A\wedge B))^{\ast})^{\circledcirc}$} \RightLabel{\hspace*{1cm}$\Pi_{4}$ and $\Pi_{5}$ are similar.}
\UnaryInfC{$((C^{\bullet}\wedge(C\ra A)^{\ast})^{\circledcirc} \wedge (C^{\bullet}\wedge(C\ra B)^{\ast})^{\circledcirc}) \ra (C^{\bullet}\wedge(C\ra (A\wedge B))^{\ast})^{\circledcirc}$} \DisplayProof

$\bs{\Pi_{4}}$\AxiomC{$C^{\bullet}\wedge(C\ra A)^{\ast}$} \RightLabel{$N$}
\UnaryInfC{$C^{\bullet}\wedge(C\ra A)^{\ast}$} \RightLabel{$N$}
\UnaryInfC{$(C\ra A)^{\ast}$} \RightLabel{$N$}
\AxiomC{$\shpos N$} \RightLabel{$M$}
\UnaryInfC{$\shpos N$} \RightLabel{$M$}
\BinaryInfC{$(C\ra A)^{\ast}$} \RightLabel{$M,\ast$}
\UnaryInfC{$C\ra A$}
\AxiomC{[$C$]} \RightLabel{$M,\ast$}
\UnaryInfC{$C$} \RightLabel{$M,\ast$}
\BinaryInfC{$A$} \RightLabel{$M,\ast$}
\AxiomC{$C^{\bullet}\wedge(C\ra B)^{\ast}$} \RightLabel{$M$}
\UnaryInfC{$C^{\bullet}\wedge(C\ra B)^{\ast}$} \RightLabel{$M$}
\UnaryInfC{$(C\ra B)^{\ast}$} \RightLabel{$M,\ast$}
\UnaryInfC{$C\ra B$} \RightLabel{$M,\ast$}
\AxiomC{[$C$]} \RightLabel{$M,\ast$}
\UnaryInfC{$C$} \RightLabel{$M,\ast$}
\BinaryInfC{$B$} \RightLabel{$M,\ast$}
\BinaryInfC{$A \wedge B$} \RightLabel{$M,\ast$}
\UnaryInfC{$C \ra (A \wedge B)$} \RightLabel{$M$}
\UnaryInfC{$(C \ra (A \wedge B))^{\ast}$}
\AxiomC{[$(C^{\bullet}\wedge(C\ra A)^{\ast})^{\circledcirc} \wedge (C^{\bullet}\wedge(C\ra B)^{\ast})^{\circledcirc}$]}
\UnaryInfC{$(C^{\bullet}\wedge(C\ra A)^{\ast})^{\circledcirc} \wedge (C^{\bullet}\wedge(C\ra B)^{\ast})^{\circledcirc}$}
\UnaryInfC{$(C^{\bullet}\wedge(C\ra A)^{\ast})^{\circledcirc}$}
\BinaryInfC{$(C^{\bullet}\wedge(C\ra A)^{\ast})^{\circledcirc}$} 
\alwaysNoLine \UnaryInfC{$\phantom{-}$} \DisplayProof

\textbf{A3}\AxiomC{$^{1}$[$(C^{\bullet}\wedge(C\ra A)^{\ast})^{\circledcirc}$]}
\UnaryInfC{$(C^{\bullet}\wedge(C\ra A)^{\ast})^{\circledcirc}$} \RightLabel{$\circledcirc$}
\UnaryInfC{$C^{\bullet}\wedge(C\ra A)^{\ast}$}
\AxiomC{$^{2}$[$(C^{\bullet}\wedge(C\ra \neg A)^{\ast})^{\circledcirc}$]}
\UnaryInfC{$(C^{\bullet}\wedge(C\ra \neg A)^{\ast})^{\circledcirc}$} \RightLabel{$\circledcirc$}
\UnaryInfC{$C^{\bullet}\wedge(C\ra \neg A)^{\ast}$}
\AxiomC{$^{3}$[$\shpos N$]} \RightLabel{$M$}
\UnaryInfC{$\Pi_{6}$}
\UnaryInfC{$\bot_{n}$}
\AxiomC{$^{3}$[$\shpos M$]} \RightLabel{$N$}
\UnaryInfC{$\Pi_{7}$}
\UnaryInfC{$\bot_{n}$} \LeftLabel{3}
\BinaryInfC{$\bot_{n}$} 
\BinaryInfC{$\bot_{n}$} \LeftLabel{2}
\UnaryInfC{$\neg (C^{\bullet}\wedge(C\ra \neg A)^{\ast})^{\circledcirc}$}
\BinaryInfC{$(C^{\bullet}\wedge(C\ra \neg A)^{\ast})^{\circledcirc}$} \LeftLabel{1}
\UnaryInfC{$(C^{\bullet}\wedge(C\ra A)^{\ast})^{\circledcirc} \ra (C^{\bullet}\wedge(C\ra \neg A)^{\ast})^{\circledcirc}$} \DisplayProof $\Pi_{6}$ and $\Pi_{7}$ similar.

$\bs{\Pi_{6}}$\AxiomC{$C^{\bullet}\wedge(C\ra \neg A)^{\ast}$} \RightLabel{$M$}
\UnaryInfC{$C^{\bullet}\wedge(C\ra \neg A)^{\ast}$} \RightLabel{$M$}
\UnaryInfC{$C^{\bullet}$} \RightLabel{$M,\bullet$}
\UnaryInfC{$C$}
\AxiomC{$^{1}$[$C$]} \RightLabel{$M,u$}
\UnaryInfC{$C$}
\AxiomC{$C^{\bullet}\wedge(C\ra A)^{\ast}$} \RightLabel{$N$}
\UnaryInfC{$C^{\bullet}\wedge(C\ra A)^{\ast}$} \RightLabel{$N$}
\UnaryInfC{$(C\ra A)^{\ast}$}
\AxiomC{$\shpos N$} \RightLabel{$M$}
\UnaryInfC{$\shpos N$} \RightLabel{$M$}
\BinaryInfC{$(C\ra A)^{\ast}$} \RightLabel{$M,\ast$}
\UnaryInfC{$C\ra A$} \RightLabel{$M,u$}
\UnaryInfC{$C\ra A$} \RightLabel{$M,u$}
\BinaryInfC{$A$} \RightLabel{$M,u$}
\AxiomC{$^{1}$[$C$]} \RightLabel{$M,u$}
\UnaryInfC{$C$}
\AxiomC{$C^{\bullet}\wedge(C\ra \neg A)^{\ast}$} \RightLabel{$M$}
\UnaryInfC{$C^{\bullet}\wedge(C\ra \neg A)^{\ast}$} \RightLabel{$M$}
\UnaryInfC{$(C\ra \neg A)^{\ast}$} \RightLabel{$M,\ast$}
\UnaryInfC{$C\ra \neg A$} \RightLabel{$M,u$}
\UnaryInfC{$C\ra \neg A$} \RightLabel{$M,u$}
\BinaryInfC{$\neg A$} \RightLabel{$M,u$}
\BinaryInfC{$\bot_{n}$}
\UnaryInfC{$\bot_{n}$} \LeftLabel{1}
\BinaryInfC{$\bot_{n}$} \DisplayProof

\textbf{A4}\AxiomC{$^{3}$[$(C^{\bullet}\wedge(C\ra \top_{n})^{\ast})^{\circledcirc}$]}
\UnaryInfC{$(C^{\bullet}\wedge(C\ra \top_{n})^{\ast})^{\circledcirc}$} \RightLabel{$\circledcirc$}
\UnaryInfC{$C^{\bullet}\wedge(C\ra \top_{n})^{\ast}$} \RightLabel{$\circledcirc$}
\UnaryInfC{$C^{\bullet}$}
\AxiomC{$^{3}$[$(C^{\bullet}\wedge(C\ra \top_{n})^{\ast})^{\circledcirc}$]}
\UnaryInfC{$(C^{\bullet}\wedge(C\ra \top_{n})^{\ast})^{\circledcirc}$}
\AxiomC{$^{2}$[$C^{\bullet}$]} \RightLabel{$N$}
\UnaryInfC{$C^{\bullet}$}
\AxiomC{$^{2}$[$C^{\bullet}$]} \RightLabel{$N$}
\UnaryInfC{$C^{\bullet}$}
\AxiomC{$^{1}$[$C$]} \RightLabel{$\circledast,\ast$}
\UnaryInfC{$C$} \LeftLabel{1} \RightLabel{$\circledast,\ast$}
\UnaryInfC{$C \ra C$} \RightLabel{$\circledast$}
\UnaryInfC{$(C \ra C)^{\ast}$} \RightLabel{$N$}
\BinaryInfC{$(C \ra C)^{\ast}$} \RightLabel{$N$}
\BinaryInfC{$C^{\bullet}\wedge(C \ra C)^{\ast}$} \RightLabel{$\circledcirc$}
\BinaryInfC{$C^{\bullet}\wedge(C \ra C)^{\ast}$}
\UnaryInfC{$(C^{\bullet}\wedge(C \ra C)^{\ast})^{\circledcirc}$} \LeftLabel{2}
\BinaryInfC{$(C^{\bullet}\wedge(C\ra C)^{\ast})^{\circledcirc}$} \LeftLabel{3}
\UnaryInfC{$(C^{\bullet}\wedge(C\ra \top_{n})^{\ast})^{\circledcirc} \ra (C^{\bullet}\wedge(C\ra C)^{\ast})^{\circledcirc}$} \DisplayProof

\textbf{A5}\AxiomC{$^{2}$[$(C^{\bullet}\wedge(C\ra \top_{n})^{\ast})^{\circledcirc}$]}
\UnaryInfC{$(C^{\bullet}\wedge(C\ra \top_{n})^{\ast})^{\circledcirc}$} \RightLabel{$\circledcirc$}
\UnaryInfC{$C^{\bullet}\wedge(C\ra \top_{n})^{\ast}$} \RightLabel{$\circledcirc$}
\UnaryInfC{$C^{\bullet}$}
\AxiomC{$^{2}$[$(C^{\bullet}\wedge(C\ra \top_{n})^{\ast})^{\circledcirc}$]}
\UnaryInfC{$(C^{\bullet}\wedge(C\ra \top_{n})^{\ast})^{\circledcirc}$}
\AxiomC{$^{1}$[$C^{\bullet}$]} \RightLabel{$N$}
\UnaryInfC{$C^{\bullet}$} \RightLabel{$N,\bullet$}
\UnaryInfC{$C$} \RightLabel{$N,\bullet$}
\UnaryInfC{$B \vee C$} \RightLabel{$N$}
\UnaryInfC{$(B \vee C)^{\bullet}$}
\AxiomC{$\phantom{-}$} \RightLabel{$N,\ast$}
\UnaryInfC{$\top_{n}$} \RightLabel{$N,\ast$}
\UnaryInfC{$(B \vee C) \ra \top_{n}$} \RightLabel{$N$}
\UnaryInfC{$((B \vee C)\ra \top_{n})^{\ast}$} \RightLabel{$N$}
\BinaryInfC{$(B \vee C))^{\bullet}\wedge((B \vee C)\ra \top_{n})^{\ast}$} \RightLabel{$\circledcirc$}
\BinaryInfC{$(B \vee C))^{\bullet}\wedge((B \vee C)\ra \top_{n})^{\ast}$}
\UnaryInfC{$((B \vee C))^{\bullet}\wedge((B \vee C)\ra \top_{n})^{\ast})^{\circledcirc}$} \LeftLabel{1}
\BinaryInfC{$((B \vee C))^{\bullet}\wedge((B \vee C)\ra \top_{n})^{\ast})^{\circledcirc}$} \LeftLabel{2}
\UnaryInfC{$(C^{\bullet}\wedge(C\ra \top_{n})^{\ast})^{\circledcirc} \ra ((B \vee C))^{\bullet}\wedge((B \vee C)\ra \top_{n})^{\ast})^{\circledcirc}$} \alwaysNoLine \UnaryInfC{$\phantom{-}$} \DisplayProof

\textbf{A6}\AxiomC{$^{1}$[$(B^{\bullet}\wedge(B\ra A)^{\ast})^{\circledcirc}\wedge(C^{\bullet}\wedge(C\ra A)^{\ast})^{\circledcirc}$]}
\UnaryInfC{$(B^{\bullet}\wedge(B\ra A)^{\ast})^{\circledcirc}$} \RightLabel{$\circledcirc$}
\UnaryInfC{$B^{\bullet}\wedge(B\ra A)^{\ast}$}
\AxiomC{$^{1}$[$(B^{\bullet}\wedge(B\ra A)^{\ast})^{\circledcirc}\wedge(C^{\bullet}\wedge(C\ra A)^{\ast})^{\circledcirc}$]}
\UnaryInfC{$(C^{\bullet}\wedge(C\ra A)^{\ast})^{\circledcirc}$} \RightLabel{$\circledcirc$}
\UnaryInfC{$C^{\bullet}\wedge(C\ra A)^{\ast}$}
\AxiomC{$\Pi_{8}$}
\UnaryInfC{$((B \vee C))^{\bullet}\wedge((B \vee C)\ra A)^{\ast})^{\circledcirc}$}
\BinaryInfC{$((B \vee C))^{\bullet}\wedge((B \vee C)\ra A)^{\ast})^{\circledcirc}$}
\BinaryInfC{$((B \vee C))^{\bullet}\wedge((B \vee C)\ra A)^{\ast})^{\circledcirc}$} \LeftLabel{1}
\UnaryInfC{$((B^{\bullet}\wedge(B\ra A)^{\ast})^{\circledcirc}\wedge(C^{\bullet}\wedge(C\ra A)^{\ast})^{\circledcirc}) \ra ((B \vee C))^{\bullet}\wedge((B \vee C)\ra A)^{\ast})^{\circledcirc}$} \alwaysNoLine \UnaryInfC{$\phantom{-}$} \DisplayProof

$\bs{\Pi_{8}}$\AxiomC{[$\shpos M$]} \RightLabel{$N$}
\UnaryInfC{$\Pi_{9}$}
\UnaryInfC{$((B \vee C))^{\bullet}\wedge((B \vee C)\ra A)^{\ast})^{\circledcirc}$}
\AxiomC{[$\shpos N$]} \RightLabel{$M$}
\UnaryInfC{$\Pi_{10}$}
\UnaryInfC{$((B \vee C))^{\bullet}\wedge((B \vee C)\ra A)^{\ast})^{\circledcirc}$}
\BinaryInfC{$((B \vee C))^{\bullet}\wedge((B \vee C)\ra A)^{\ast})^{\circledcirc}$} \alwaysNoLine \UnaryInfC{$\phantom{-}$} \DisplayProof $\phantom{---}\Pi_{9}$ and $\Pi_{10}$ similar.

$\bs{\Pi_{9}}$\AxiomC{$(B^{\bullet}\wedge(B\ra A)^{\ast})^{\circledcirc}\wedge(C^{\bullet}\wedge(C\ra A)^{\ast})^{\circledcirc}$}
\UnaryInfC{$(B^{\bullet}\wedge(B\ra A)^{\ast})^{\circledcirc}$}
\AxiomC{$^{1}$[$B^{\bullet}\wedge(B\ra A)^{\ast}$]} \RightLabel{$N$}
\UnaryInfC{$B^{\bullet}\wedge(B\ra A)^{\ast}$} \RightLabel{$N$}
\UnaryInfC{$B^{\bullet}$} \RightLabel{$N,\bullet$}
\UnaryInfC{$B$} \RightLabel{$N,\bullet$}
\UnaryInfC{$B \vee C$} \RightLabel{$N$}
\UnaryInfC{$(B \vee C)^{\bullet}$}
\AxiomC{$\shpos M$} \RightLabel{$N$}
\UnaryInfC{$\Pi_{11}$} \RightLabel{$N$}
\UnaryInfC{$((B \vee C)\ra A)^{\ast}$} \RightLabel{$N$}
\BinaryInfC{$(B \vee C))^{\bullet}\wedge((B \vee C)\ra A)^{\ast}$} \LeftLabel{1} \RightLabel{$\circledcirc$}
\BinaryInfC{$(B \vee C))^{\bullet}\wedge((B \vee C)\ra A)^{\ast}$}
\UnaryInfC{$((B \vee C))^{\bullet}\wedge((B \vee C)\ra A)^{\ast})^{\circledcirc}$} \DisplayProof

$\bs{\Pi_{11}}$ \AxiomC{$^{2}$[$B \vee C$]} \RightLabel{$N,\ast$}
\UnaryInfC{$B \vee C$}
\AxiomC{$^{3}$[$B$]} \RightLabel{$N,\ast$}
\UnaryInfC{$B$}
\AxiomC{$B^{\bullet}\wedge(B\ra A)^{\ast}$} \RightLabel{$N$}
\UnaryInfC{$B^{\bullet}\wedge(B\ra A)^{\ast}$} \RightLabel{$N$}
\UnaryInfC{$(B\ra A)^{\ast}$} \RightLabel{$N,\ast$}
\UnaryInfC{$B\ra A$} \RightLabel{$N,\ast$}
\BinaryInfC{$A$}
\AxiomC{$^{3}$[$C$]} \RightLabel{$N,\ast$}
\UnaryInfC{$C$}
\AxiomC{$C^{\bullet}\wedge(C\ra A)^{\ast}$} \RightLabel{$M$}
\UnaryInfC{$C^{\bullet}\wedge(C\ra A)^{\ast}$} \RightLabel{$M$}
\UnaryInfC{$(C\ra A)^{\ast}$} \RightLabel{$M$}
\AxiomC{$\shpos M$} \RightLabel{$N$}
\UnaryInfC{$\shpos M$} \RightLabel{$N$}
\BinaryInfC{$(C\ra A)^{\ast}$} \RightLabel{$N,\ast$}
\UnaryInfC{$C\ra A$} \RightLabel{$N,\ast$}
\BinaryInfC{$A$} \RightLabel{$N,\ast$}
\BinaryInfC{$A$} \LeftLabel{3} \RightLabel{$N,\ast$}
\BinaryInfC{$A$} \LeftLabel{2} \RightLabel{$N,\ast$}
\UnaryInfC{$(B \vee C)\ra A$} \RightLabel{$N$}
\UnaryInfC{$((B \vee C)\ra A)^{\ast}$}
\alwaysNoLine \UnaryInfC{$\phantom{-}$}\DisplayProof

\textbf{A7}\AxiomC{$^{1}$[$\neg ((C^{\bullet}\wedge(C \ra \neg \bot_{n})^{\ast})^{\circledcirc}) \wedge ( (B \vee C )^{\bullet} \wedge ((B \vee C) \ra A)^{\ast})^{\circledcirc}$]}
\UnaryInfC{$\neg ((C^{\bullet}\wedge(C \ra \neg \bot_{n})^{\ast})^{\circledcirc}) \wedge ( (B \vee C )^{\bullet} \wedge ((B \vee C) \ra A)^{\ast})^{\circledcirc}$}
\UnaryInfC{$( (B \vee C )^{\bullet} \wedge ((B \vee C) \ra A)^{\ast})^{\circledcirc}$} \RightLabel{$\circledcirc$}
\UnaryInfC{$(B \vee C )^{\bullet} \wedge ((B \vee C) \ra A)^{\ast}$}
\AxiomC{$\Pi_{12}$}
\UnaryInfC{$(B^{\bullet} \wedge (B \ra A)^{\ast})^{\circledcirc}$}
\BinaryInfC{$(B^{\bullet} \wedge (B \ra A)^{\ast})^{\circledcirc}$} \LeftLabel{1}
\UnaryInfC{$(\neg ((C^{\bullet}\wedge(C \ra \neg \bot_{n})^{\ast})^{\circledcirc}) \wedge ( (B \vee C )^{\bullet} \wedge ((B \vee C) \ra A)^{\ast})^{\circledcirc}) \ra (B^{\bullet} \wedge (B \ra A)^{\ast})^{\circledcirc}$} \alwaysNoLine \UnaryInfC{$\phantom{-}$} \DisplayProof

$\bs{\Pi_{12}}$\AxiomC{$(B \vee C )^{\bullet} \wedge ((B \vee C) \ra A)^{\ast}$} \RightLabel{$N$}
\UnaryInfC{$(B \vee C )^{\bullet} \wedge ((B \vee C) \ra A)^{\ast}$} \RightLabel{$N$}
\UnaryInfC{$(B \vee C )^{\bullet}$} \RightLabel{$N,\bullet$}
\UnaryInfC{$B \vee C$}
\AxiomC{$^{1}$[$B$]} \RightLabel{$N,\bullet$}
\UnaryInfC{$B$} \RightLabel{$N$}
\UnaryInfC{$\Pi_{13}$}
\UnaryInfC{$(B^{\bullet} \wedge (B \ra A)^{\ast})^{\circledcirc}$}
\AxiomC{$^{1}$[$C$]} \RightLabel{$N,\bullet$}
\UnaryInfC{$C$} \RightLabel{$N$}
\UnaryInfC{$\Pi_{14}$}
\UnaryInfC{$(B^{\bullet} \wedge (B \ra A)^{\ast})^{\circledcirc}$} \LeftLabel{1}
\TrinaryInfC{$(B^{\bullet} \wedge (B \ra A)^{\ast})^{\circledcirc}$} \alwaysNoLine \UnaryInfC{$\phantom{-}$} \DisplayProof

$\bs{\Pi_{13}}$\AxiomC{$\neg ((C^{\bullet}\wedge(C \ra \neg \bot_{n})^{\ast})^{\circledcirc}) \wedge ( (B \vee C )^{\bullet} \wedge ((B \vee C) \ra A)^{\ast})^{\circledcirc}$}
\UnaryInfC{$\neg ((C^{\bullet}\wedge(C \ra \neg \bot_{n})^{\ast})^{\circledcirc}) \wedge ( (B \vee C )^{\bullet} \wedge ((B \vee C) \ra A)^{\ast})^{\circledcirc}$}
\UnaryInfC{$( (B \vee C )^{\bullet} \wedge ((B \vee C) \ra A)^{\ast})^{\circledcirc}$}
\AxiomC{$B$} \RightLabel{$N,\bullet$}
\UnaryInfC{$B$} \RightLabel{$N$}
\UnaryInfC{$B^{\bullet}$}
\AxiomC{$^{1}$[$B$]} \RightLabel{$N,\ast$}
\UnaryInfC{$B$} \RightLabel{$N,\ast$}
\UnaryInfC{$B \vee C$}
\AxiomC{$(B \vee C )^{\bullet} \wedge ((B \vee C) \ra A)^{\ast}$} \RightLabel{$N$}
\UnaryInfC{$(B \vee C )^{\bullet} \wedge ((B \vee C) \ra A)^{\ast}$} \RightLabel{$N$}
\UnaryInfC{$((B \vee C) \ra A)^{\ast}$} \RightLabel{$N,\ast$}
\UnaryInfC{$(B \vee C) \ra A$} \RightLabel{$N,\ast$}
\BinaryInfC{$A$} \LeftLabel{1} \RightLabel{$N,\ast$}
\UnaryInfC{$B \ra A$} \RightLabel{$N$}
\UnaryInfC{$(B \ra A)^{\ast}$} \RightLabel{$N$}
\BinaryInfC{$B^{\bullet} \wedge (B \ra A)^{\ast}$} \RightLabel{$\circledcirc$}
\BinaryInfC{$B^{\bullet} \wedge (B \ra A)^{\ast}$}
\UnaryInfC{$(B^{\bullet} \wedge (B \ra A)^{\ast})^{\circledcirc}$}
\alwaysNoLine \UnaryInfC{$\phantom{-}$}\DisplayProof

$\bs{\Pi_{14}}$\AxiomC{$\neg ((C^{\bullet}\wedge(C \ra \neg \bot_{n})^{\ast})^{\circledcirc}) \wedge ( (B \vee C )^{\bullet} \wedge ((B \vee C) \ra A)^{\ast})^{\circledcirc}$}
\UnaryInfC{$\neg ((C^{\bullet}\wedge(C \ra \neg \bot_{n})^{\ast})^{\circledcirc}) \wedge ( (B \vee C )^{\bullet} \wedge ((B \vee C) \ra A)^{\ast})^{\circledcirc}$}
\UnaryInfC{$\neg ((C^{\bullet}\wedge(C \ra \neg \bot_{n})^{\ast})^{\circledcirc})$}
\AxiomC{$C$} \RightLabel{$N,\bullet$}
\UnaryInfC{$C$}
\UnaryInfC{$\Pi_{15}$}
\UnaryInfC{$(C^{\bullet}\wedge(C \ra \neg \bot_{n})^{\ast})^{\circledcirc}$} \RightLabel{$N,\ast$}
\BinaryInfC{$\bot_{n}$}
\UnaryInfC{$(B^{\bullet} \wedge (B \ra A)^{\ast})^{\circledcirc}$} \DisplayProof

$\bs{\Pi_{15}}$\AxiomC{$\neg ((C^{\bullet}\wedge(C \ra \neg \bot_{n})^{\ast})^{\circledcirc}) \wedge ( (B \vee C )^{\bullet} \wedge ((B \vee C) \ra A)^{\ast})^{\circledcirc}$}
\UnaryInfC{$\neg ((C^{\bullet}\wedge(C \ra \neg \bot_{n})^{\ast})^{\circledcirc}) \wedge ( (B \vee C )^{\bullet} \wedge ((B \vee C) \ra A)^{\ast})^{\circledcirc}$}
\UnaryInfC{$( (B \vee C )^{\bullet} \wedge ((B \vee C) \ra A)^{\ast})^{\circledcirc}$}
\AxiomC{$C$} \RightLabel{$N,\bullet$}
\UnaryInfC{$C$} \RightLabel{$N$}
\UnaryInfC{$C^{\bullet}$} \RightLabel{$N$}
\AxiomC{[$\bot_{n}$]} \RightLabel{$N,\ast$}
\UnaryInfC{$\bot_{n}$} \RightLabel{$N,\ast$}
\UnaryInfC{$\neg \bot_{n}$} \RightLabel{$N,\ast$}
\UnaryInfC{$C \ra \neg \bot_{n}$} \RightLabel{$N$}
\UnaryInfC{$(C \ra \neg \bot_{n})^{\ast}$} \RightLabel{$N$}
\BinaryInfC{$C^{\bullet}\wedge(C \ra \neg \bot_{n})^{\ast}$} \RightLabel{$\circledcirc$}
\BinaryInfC{$C^{\bullet}\wedge(C \ra \neg \bot_{n})^{\ast}$}
\UnaryInfC{$(C^{\bullet}\wedge(C \ra \neg \bot_{n})^{\ast})^{\circledcirc}$} \alwaysNoLine \UnaryInfC{$\phantom{-}$} \DisplayProof

\textbf{A8}\AxiomC{$^{1}$[$\neg (((B \vee C)^{\bullet}\wedge((B \vee C) \ra \neg B)^{\ast})^{\circledcirc}) \wedge ( (B \vee C )^{\bullet} \wedge ((B \vee C) \ra A)^{\ast})^{\circledcirc}$]}
\UnaryInfC{$\neg (((B \vee C)^{\bullet}\wedge((B \vee C) \ra \neg B)^{\ast})^{\circledcirc}) \wedge ( (B \vee C )^{\bullet} \wedge ((B \vee C) \ra A)^{\ast})^{\circledcirc}$}
\UnaryInfC{$( (B \vee C )^{\bullet} \wedge ((B \vee C) \ra A)^{\ast})^{\circledcirc}$} \RightLabel{$\circledcirc$}
\UnaryInfC{$(B \vee C )^{\bullet} \wedge ((B \vee C) \ra A)^{\ast}$}
\AxiomC{$\Pi_{16}$}
\UnaryInfC{$(B^{\bullet} \wedge (B \ra A)^{\ast})^{\circledcirc}$}
\BinaryInfC{$(B^{\bullet} \wedge (B \ra A)^{\ast})^{\circledcirc}$} \LeftLabel{1}
\UnaryInfC{$(\neg (((B \vee C)^{\bullet}\wedge((B \vee C) \ra \neg B)^{\ast})^{\circledcirc}) \wedge ( (B \vee C )^{\bullet} \wedge ((B \vee C) \ra A)^{\ast})^{\circledcirc}) \ra (B^{\bullet} \wedge (B \ra A)^{\ast})^{\circledcirc}$} \alwaysNoLine \UnaryInfC{$\phantom{-}$} \DisplayProof

$\bs{\Pi_{16}}$\AxiomC{$\neg (((B \vee C)^{\bullet}\wedge((B \vee C) \ra \neg B)^{\ast})^{\circledcirc}) \wedge ( (B \vee C )^{\bullet} \wedge ((B \vee C) \ra A)^{\ast})^{\circledcirc}$}
\UnaryInfC{$\neg (((B \vee C)^{\bullet}\wedge((B \vee C) \ra \neg B)^{\ast})^{\circledcirc}) \wedge ( (B \vee C )^{\bullet} \wedge ((B \vee C) \ra A)^{\ast})^{\circledcirc}$}
\UnaryInfC{$( (B \vee C )^{\bullet} \wedge ((B \vee C) \ra A)^{\ast})^{\circledcirc}$}
\AxiomC{$\Pi_{17}$} \RightLabel{$N$}
\UnaryInfC{$B^{\bullet} \wedge (B \ra A)^{\ast}$} \RightLabel{$\circledcirc$}
\BinaryInfC{$B^{\bullet} \wedge (B \ra A)^{\ast}$} \alwaysNoLine \UnaryInfC{$\phantom{-}$} \DisplayProof

$\bs{\Pi_{17}}$\AxiomC{$\Pi_{18}$} \RightLabel{$N$}
\UnaryInfC{$B^{\bullet} \vee \neg (B^{\bullet})$} \RightLabel{$N$}
\AxiomC{[$B^{\bullet}$]} \RightLabel{$N$}
\UnaryInfC{$\Pi_{19}$} \RightLabel{$N$}
\UnaryInfC{$B^{\bullet} \wedge (B \ra A)^{\ast}$} \RightLabel{$N$}
\AxiomC{[$\neg(B^{\bullet})$]} \RightLabel{$N$}
\UnaryInfC{$\Pi_{20}$} \RightLabel{$N$}
\UnaryInfC{$B^{\bullet} \wedge (B \ra A)^{\ast}$} \RightLabel{$N$}
\TrinaryInfC{$B^{\bullet} \wedge (B \ra A)^{\ast}$} \alwaysNoLine \UnaryInfC{$\phantom{-}$} \DisplayProof

$\bs{\Pi_{18}}$\AxiomC{$^{2}$[$\neg(B^{\bullet} \vee \neg (B^{\bullet}))$]} \RightLabel{$N$}
\UnaryInfC{$B^{\bullet} \vee \neg (B^{\bullet})$} \RightLabel{$N$}
\AxiomC{$^{2}$[$\neg(B^{\bullet} \vee \neg (B^{\bullet}))$]} \RightLabel{$N$}
\UnaryInfC{$B^{\bullet} \vee \neg (B^{\bullet})$} \RightLabel{$N$}
\AxiomC{$^{1}$[$B^{\bullet}$]}
\UnaryInfC{$B^{\bullet}$} \RightLabel{$N$}
\UnaryInfC{$B^{\bullet} \vee \neg (B^{\bullet})$} \RightLabel{$N$}
\BinaryInfC{$\bot_{w}$} \LeftLabel{1} \RightLabel{$N$}
\UnaryInfC{$\neg (B^{\bullet})$} \RightLabel{$N$}
\UnaryInfC{$B^{\bullet} \vee \neg (B^{\bullet})$} \RightLabel{$N$}
\BinaryInfC{$\bot_{w}$} \LeftLabel{2} \RightLabel{$N$}
\UnaryInfC{$B^{\bullet} \vee \neg (B^{\bullet})$} \alwaysNoLine \UnaryInfC{$\phantom{-}$} \DisplayProof

$\bs{\Pi_{19}}$\AxiomC{$B^{\bullet}$} \RightLabel{$N$}
\UnaryInfC{$B^{\bullet}$} \RightLabel{$N$}
\AxiomC{[$B$]} \RightLabel{$N,\ast$}
\UnaryInfC{$B$} \RightLabel{$N,\ast$}
\UnaryInfC{$B \vee C$} \RightLabel{$N,\ast$}
\AxiomC{$(B \vee C )^{\bullet} \wedge ((B \vee C) \ra A)^{\ast}$} \RightLabel{$N$}
\UnaryInfC{$(B \vee C )^{\bullet} \wedge ((B \vee C) \ra A)^{\ast}$} \RightLabel{$N$}
\UnaryInfC{$((B \vee C) \ra A)^{\ast}$} \RightLabel{$N,\ast$}
\UnaryInfC{$(B \vee C)\ra A$} \RightLabel{$N,\ast$}
\BinaryInfC{$A$} \RightLabel{$N,\ast$}
\UnaryInfC{$B \ra A$} \RightLabel{$N$}
\UnaryInfC{$(B \ra A)^{\ast}$} \RightLabel{$N$}
\BinaryInfC{$B^{\bullet} \wedge (B \ra A)^{\ast}$} \alwaysNoLine \UnaryInfC{$\phantom{-}$} \DisplayProof

$\bs{\Pi_{20}}$\AxiomC{$\neg (((B \vee C)^{\bullet}\wedge((B \vee C) \ra \neg B)^{\ast})^{\circledcirc}) \wedge ( (B \vee C )^{\bullet} \wedge ((B \vee C) \ra A)^{\ast})^{\circledcirc}$}
\UnaryInfC{$\neg (((B \vee C)^{\bullet}\wedge((B \vee C) \ra \neg B)^{\ast})^{\circledcirc}) \wedge ( (B \vee C )^{\bullet} \wedge ((B \vee C) \ra A)^{\ast})^{\circledcirc}$}
\UnaryInfC{$( (B \vee C )^{\bullet} \wedge ((B \vee C) \ra A)^{\ast})^{\circledcirc}$}
\AxiomC{$(B \vee C )^{\bullet} \wedge ((B \vee C) \ra A)^{\ast}$} \RightLabel{$N$}
\UnaryInfC{$(B \vee C )^{\bullet} \wedge ((B \vee C) \ra A)^{\ast}$} \RightLabel{$N$}
\UnaryInfC{$(B \vee C )^{\bullet}$} \RightLabel{$N$}
\AxiomC{$\neg(B^{\bullet})$} \RightLabel{$N$}
\UnaryInfC{$\neg(B^{\bullet})$} \RightLabel{$N$}
\AxiomC{[$B$]} \RightLabel{$N,u$}
\UnaryInfC{$B$} \RightLabel{$N,\bullet$}
\UnaryInfC{$B$} \RightLabel{$N$}
\UnaryInfC{$B^{\bullet}$} \RightLabel{$N$}
\BinaryInfC{$\bot_{n}^{u}$} \RightLabel{$N,u$}
\UnaryInfC{$\bot_{n}$} \RightLabel{$N,u$}
\UnaryInfC{$\neg B$} \RightLabel{$N,\ast$}
\UnaryInfC{$\neg B$} \RightLabel{$N,\ast$}
\UnaryInfC{$(B \vee C) \ra \neg B$} \RightLabel{$N$}
\UnaryInfC{$((B \vee C) \ra \neg B)^{\ast}$} \RightLabel{$N$}
\BinaryInfC{$(B \vee C )^{\bullet} \wedge ((B \vee C) \ra \neg B)^{\ast}$} \RightLabel{$N$}
\UnaryInfC{$(B \vee C )^{\bullet} \wedge ((B \vee C) \ra \neg B)^{\ast}$} \RightLabel{$\circledcirc$}
\BinaryInfC{$(B \vee C )^{\bullet} \wedge ((B \vee C) \ra \neg B)^{\ast}$}
\UnaryInfC{$\Pi_{21}$} \alwaysNoLine \UnaryInfC{$\phantom{-}$} \DisplayProof\end{center}\end{landscape}

\begin{center}
\begin{small}
$\bs{\Pi_{21}}$\AxiomC{$\neg (((B \vee C)^{\bullet}\wedge((B \vee C) \ra \neg B)^{\ast})^{\circledcirc}) \wedge ( (B \vee C )^{\bullet} \wedge ((B \vee C) \ra A)^{\ast})^{\circledcirc}$}
\UnaryInfC{$\neg (((B \vee C)^{\bullet}\wedge((B \vee C) \ra \neg B)^{\ast})^{\circledcirc}) \wedge ( (B \vee C )^{\bullet} \wedge ((B \vee C) \ra A)^{\ast})^{\circledcirc}$}
\UnaryInfC{$\neg (((B \vee C)^{\bullet}\wedge((B \vee C) \ra \neg B)^{\ast})^{\circledcirc})$}
\AxiomC{$\phantom{-}$} \RightLabel{$\circledcirc$}
\UnaryInfC{$(B \vee C )^{\bullet} \wedge ((B \vee C) \ra \neg B)^{\ast}$}
\UnaryInfC{$((B \vee C )^{\bullet} \wedge ((B \vee C) \ra \neg B)^{\ast})^{\circledcirc}$}
\BinaryInfC{$\bot_{n}$}
\UnaryInfC{$(B^{\bullet} \wedge (B \ra A)^{\ast})^{N}$} \RightLabel{$N$}
\UnaryInfC{$B^{\bullet} \wedge (B \ra A)^{\ast}$} \alwaysNoLine \UnaryInfC{$\phantom{-}$} \DisplayProof\end{small}\end{center}

\section*{Conclusions}

We could represent the rules and axioms of the very basic $\bs{CO}$ logic and prove its relative completeness in PUC-Logic framework. It means that every deontic logic proposed by Lewis can be represented and has its completeness in PUC-Logic.


\section*{References}
\begin{thebibliography}{1}
\newcommand{\enquote}[1]{``#1''}

\bibitem{Lewis}
Lewis, D. K., \enquote{Counterfactuals}, Blackwell Publishing, 2008.

\bibitem{LewisPapers}
Lewis, D. K., \enquote{Papers in ethics and social philosophy}, Cambridge University Press, 2000.

\bibitem{Goodman}
Goodman, N., \enquote{Fact, Fiction, and Forecast}, 4th Edition, Harvard University Press, 1983.

\bibitem{Bell}
Bell, J.~L., \enquote{Toposes and Local Set Theories}, Dover Publications,
  2008.
  
\bibitem{Knuth}
  Knuth, D. E., \enquote{Semantics of context-free languages}, Mathematical Systems
Theory 2 (1968).

\bibitem{Goldblatt}
Goldblatt, R., \enquote{Topoi: The categorical analysis of logic}, Dover, 2006.

\bibitem{Goldblatt2}
Goldblatt, R., \enquote{Logics of time and computation}, CSLI lecture notes, 1992.

\bibitem{Prawitz}
Prawitz, D., \enquote{Natural Deduction: a proof-theoretical study}, Dover, 2006.

\bibitem{Naufel}
do~Amaral, F.~N. and E.~H. Haeusler, \enquote{Using the internal logic of a topos
  to model search spaces for problems}, Logic Journal of IGPL  (2007).

\bibitem{Hermann}
Menezes, P.~B. and E.~H. Haeusler, \enquote{Teoria das Categorias para Ci\^{e}ncia da Computa\c{c}\~{a}o}, Editora Sagra Luzatto, 2006.

\bibitem{Ramsey}
Ramsey, F.~P., \enquote{Philosophical papers}, Cambridge University Press,
  1990.

\bibitem{Gent} 
Gent, I. P., \enquote{A Sequent- or Tableau-style System for Lewis's Counterfactual Logic VC}, Notre Dame Journal of Formal Logic, vol. 33, no. 3, pp. 369-382, 1992.

\bibitem{Bonevac}
Bonevac, D., \enquote{Deduction: Introductory Symbolic Logic}, Blackwell, 2003.

\bibitem{Sano}
Sano, K., \enquote{Hybrid counterfactual logics}, Journal of Logic, Language and Information, volume 18, No. 4, pp 515-539, 2009.

\bibitem{Escobar}
L\'{o}pez-Escobar, E.G.K., \enquote{Implicational Logics in Natural Deduction Systems}, Journal of Symbolic Logic, Vol. 47, No. 1, pp. 184-186, 1982

\bibitem{PUC-Logic-ArXiv}
Fernandes, R.Q.A., Haeusler, E.H., Pereira, L.C.P.D., \enquote{PUC-Logic}, available at http://arxiv.org/abs/1402.1535

\bibitem{Fernandes}
Fernandes, R.Q.A., Haeusler, E.H., Pereira, L.C.P.D., \enquote{A Natural Deduction System for Counterfactual Logic}, in XVI Encontro Brasileiro de L\'{o}gica, Petr\'{o}polis, 2011.

\bibitem{LSFA09}
Fernandes, R.Q.A., Haeusler, E.H., \enquote{A Topos-Theoretic Approach to Counterfactual Logic}, in Fourth Workshop on Logical and Semantic Frameworks, Bras\'{i}lia, 2009. Pre-proceedings, 2009.

\bibitem{Hansson}
Hansson, B., \enquote{An Analysis of some Deontic Logics}, No\^{u}s, Vol. 3, No. 4, pp. 373-398, 1969.

\bibitem{vanDalen}
van Dalen, D., \enquote{Logic and Structure}, Springer, 2008.

\bibitem{Libkin}
Libkin, L., \enquote{Elements of Finite Model Theory}, Springer, 2010.

\bibitem{Troelstra}
Troelstra, A. S., Schwichtenberg, H., \enquote{Basic Proof Theory}, Cambridge University Press, 2000.

\bibitem{Lambert}
Lambert, K., \enquote{Free Logic: selected essays}, Cambridge University Press, 2004.

\bibitem{Cook}
Cook, S. A., \enquote{The complexity of theorem proving procedures}, In 3rd Annual ACM
Symposium on the Theory of Computation, pages 151-158, 1971.

\bibitem{Statman}
Statman, R., \enquote{Intuitioinistic propositional logic is polinomial-space complete}, Journal of Theoretical Computer Science, vol. 9, no. 1, pp. 67-72, 1979.

\bibitem{Jelia2012}
Lellmann, B., Pattinson, D., \enquote{Sequent Systems for Lewis' Conditional Logics}, In 13th European Conference on Logics in Artificial Intelligence, 2012.

\end{thebibliography}
\end{document}